\documentclass[12pt]{article}

\usepackage[english]{babel}

\usepackage[letterpaper,top=2cm,bottom=2cm,left=3cm,right=3cm,marginparwidth=1.75cm]{geometry}

\usepackage{amsmath}
\usepackage{graphicx}
\usepackage[colorlinks=true, allcolors=blue]{hyperref}

\usepackage{multirow}
\usepackage{tabularx}
\usepackage{subcaption}
\usepackage{comment}
\usepackage{bm}
\usepackage{algorithm2e}
\usepackage{authblk}
\usepackage{adjustbox}
\usepackage{makecell}

\usepackage{amssymb}
\usepackage{pdflscape}

\usepackage{color}
\usepackage{colortbl}
\usepackage[table]{xcolor}

\newcommand{\absdiv}[1]{%
  \par\addvspace{.5\baselineskip}
  \noindent\textbf{#1}\quad\ignorespaces
}

\usepackage{titlesec}
\usepackage{hyperref}

\titleclass{\subsubsubsection}{straight}[\subsection]

\newcounter{subsubsubsection}[subsubsection]
\renewcommand\thesubsubsubsection{\thesubsubsection.\arabic{subsubsubsection}}

\titleformat{\subsubsubsection}
  {\normalfont\normalsize\bfseries}{\thesubsubsubsection}{1em}{}
\titlespacing*{\subsubsubsection}
{0pt}{3.25ex plus 1ex minus .2ex}{1.5ex plus .2ex}

\makeatletter
\renewcommand\paragraph{\@startsection{paragraph}{5}{15pt}%
  {3.25ex \@plus1ex \@minus.2ex}%
  {-0.5em}%
  {\normalfont\normalsize\bfseries}}
\renewcommand\subparagraph{\@startsection{subparagraph}{6}{\parindent}%
  {3.25ex \@plus1ex \@minus .2ex}%
  {-1em}%
  {\normalfont\normalsize\bfseries}}
\def\toclevel@subsubsubsection{4}
\def\toclevel@paragraph{5}
\def\toclevel@subparagraph{6}
\def\l@subsubsubsection{\@dottedtocline{4}{7em}{4em}}
\def\l@paragraph{\@dottedtocline{5}{10em}{5em}}
\def\l@subparagraph{\@dottedtocline{6}{14em}{6em}}
\makeatother

\newtheorem{example}{Example}

\usepackage{tcolorbox}

\newtcolorbox{highlightsbox}{
    colback=yellow!10, 
    colframe=yellow!50!black, 
    coltitle=black, 
    fonttitle=\bfseries, 
    title=Highlights, 
    sharp corners, 
    boxrule=0.5mm 
}

\newcommand{\killpunct}[1]{}

\usepackage[square,numbers,sort]{natbib}
\usepackage{doi}

\title{A decision analysis model for colorectal cancer screening}

\author[a]{Daniel Corrales\thanks{Corresponding author.} }
\author[a]{David Ríos Insua}
\author[b]{Marino J. González}

\affil[a]{Institute of Mathematical Sciences, ICMAT-CSIC, 28049 Madrid, Spain}
\affil[b]{Department of Economics, University of La Rioja, 26006 Logroño, Spain}
\date{}

\begin{document}

\maketitle

\begin{abstract}
\absdiv{Background and Objective} With minor differences, most national colorectal cancer (CRC) screening programs in Europe consist of one-size-fits-all aged-based strategies. This paper provides a decision analysis-based approach to personalized CRC screening, supporting decisions concerning whether and which screening method to consider and/or whether a colonoscopy should be administered.

\absdiv{Methods} We use an influence diagram which characterizes CRC risk with respect to different variables of interest and includes comfort, costs, complications, and information as decision criteria, the last one assessed through information theory measures. The criteria are integrated with a multi-attribute utility model. Optimal screening policies are then computed.

\absdiv{Results}
The proposed model is used to support personalized individual screening based on relevant characteristics. It serves to assess existing national screening programs and design new ones. In particular, it suggests replacing current age-based strategies followed in many European countries by more personalized strategies based on the type of model proposed. Additionally, the model facilitates benchmarking of novel screening devices.

\absdiv{Conclusions}
This work creates a framework supporting personalized CRC screening 
  improving upon current age-based screening strategies.
\end{abstract}

\section{Introduction}

Although colorectal cancer (CRC) is the third most common type of cancer worldwide, making up for about 10 \% of all cases, only about 14 \% of susceptible EU citizens participate in CRC screening programs. 
 At the moment, these are mainly {\em one-size-fits-all } strategies \citep{KASTRINOS} focused on age and using fecal testing and colonoscopy (CS). The latter is considered highly invasive, influencing quite negatively program uptake among the general population \cite{ferrari2021towards}. 
 It has been reported \cite{ola2024utilization} through the European Health Interview Survey (2018-2020) that the coverage of faecal tests in the population aged 50-74 varies according to the organisation of the screening programme. In countries with fully rolled-out programmes it varies from 37.7\% in Croatia to 74.9\% in Denmark; in turn, in countries without programmes or with localised programmes it varies from 6.3\% in Bulgaria to 34.2\% in Latvia. This study also found statistical evidence that less participation in screening programmes corresponds to unmarried adult patients which have a low educational level, live in rural areas, have an unhealthy lifestyle or spend long periods without health checks. 
 Hence, on the one hand, there is a clear need for accurate, non-invasive, cost-effective screening tests using novel technologies as well as, on the other, for raising awareness about the disease and its 
 early detection. In addition, genetic, socioeconomic, and behavioral factors can influence the development of CRC and lead to different disease onsets. Recent studies \cite{sung2024colorectal} show that early-onset CRC incidence is rising in several countries, pointing at some of the mentioned risk factors as potential causes. Personalized screening strategies that consider these factors seem relevant together with decision 
     tools with guarantees that support experts in their implementation.
 This is especially important given the discrepancies among the strategies implemented in different countries \citep{barre2020cost}.

As an example, in 2013 the European Union (EU) drafted guidelines for quality assurance in CRC screening and diagnosis, recommending the use of national programs based on fecal immunochemical tests (FIT) and CS \cite{european2013european}. Such guidelines
 suggested that the identification and invitation of the target population, diagnosis and management of the disease, and the appropriate surveillance of people with detected lesions could be achieved by following and adopting the proposed recommendations. Interestingly, though the screening program structure is similar across many countries, details such as age or test cutoffs differ \citep{ferlizza2021roadmap}. In Western Europe, most programs are regional or national and based on FIT or gFOBT (guiac fecal occult blood test), 
 with wide differences in participation rates. In Eastern Europe, countries mostly rely on pilot or opportunistic programs, also based on FIT or gFOBT, with lower participation rates.

 Relevant analyses have been developed to assess the importance of various factors influencing the effectiveness of CRC screening programs. A significant portion of the work in the field is dedicated to applying Markov models for cost-effectiveness analyses of screening \cite{fouladi2024cost}, or to modeling patients' preferences concerning screening methods taking into account their risk tolerance \cite{taksler2017modeling}. An important and complementary approach to further improve the evaluation of program needs, effectiveness, and objectives is the development of CRC predictive models that form the foundation of decision-support tools for screening. In particular, Bayesian Networks (BNs) \citep{Jensen1996} are used to infer the influence that certain risk factors can have on CRC and how they could affect the decisions made by policymakers. Influence Diagrams (IDs) \citep{shachter} extend BNs by incorporating decision nodes and utility variables to support multiple criteria decisions under uncertainty. These analyses have proved to be relevant in medical decision-making \cite{roberts2012conceptualizing, dolan2014can}.

This paper constructs a decision support model that serves to benchmark CRC screening programs, among other uses. We draw upon earlier work on CRC risk assessment through BNs \citep{corrales2024colorectal} 
 and complement it to design an ID that identifies the screening methods adopted, their impacts, and their associated utilities. It will be used to discuss relevant policy questions concerning CRC screening. For this, Section \ref{section:2} defines how the problem is structured and present the underlying prediction and preference models. After that, Section \ref{section:3} describes a set of important use cases including supporting personalized screening decisions, assessing national screening strategies, designing such strategies, and specifying benchmarks for novel screening technologies. Software supporting the proposed approach produced in GeNIe \cite{bayesfusion2024genie} and its Python wrapper PySMILE is available at 
\cite{Corrales-Decision-Model-Screening-CRC-2024} 
for reproducibility purposes.

\section{Methods}\label{section:2}

\subsection{CRC screening decision problem structure}
This section describes the structure of our CRC personalized screening decision support model. 

\subsubsection{CRC underlying predictive model}\label{section:2.1}
\begin{figure}[ht]
\centering
\includegraphics[width=0.8\textwidth]{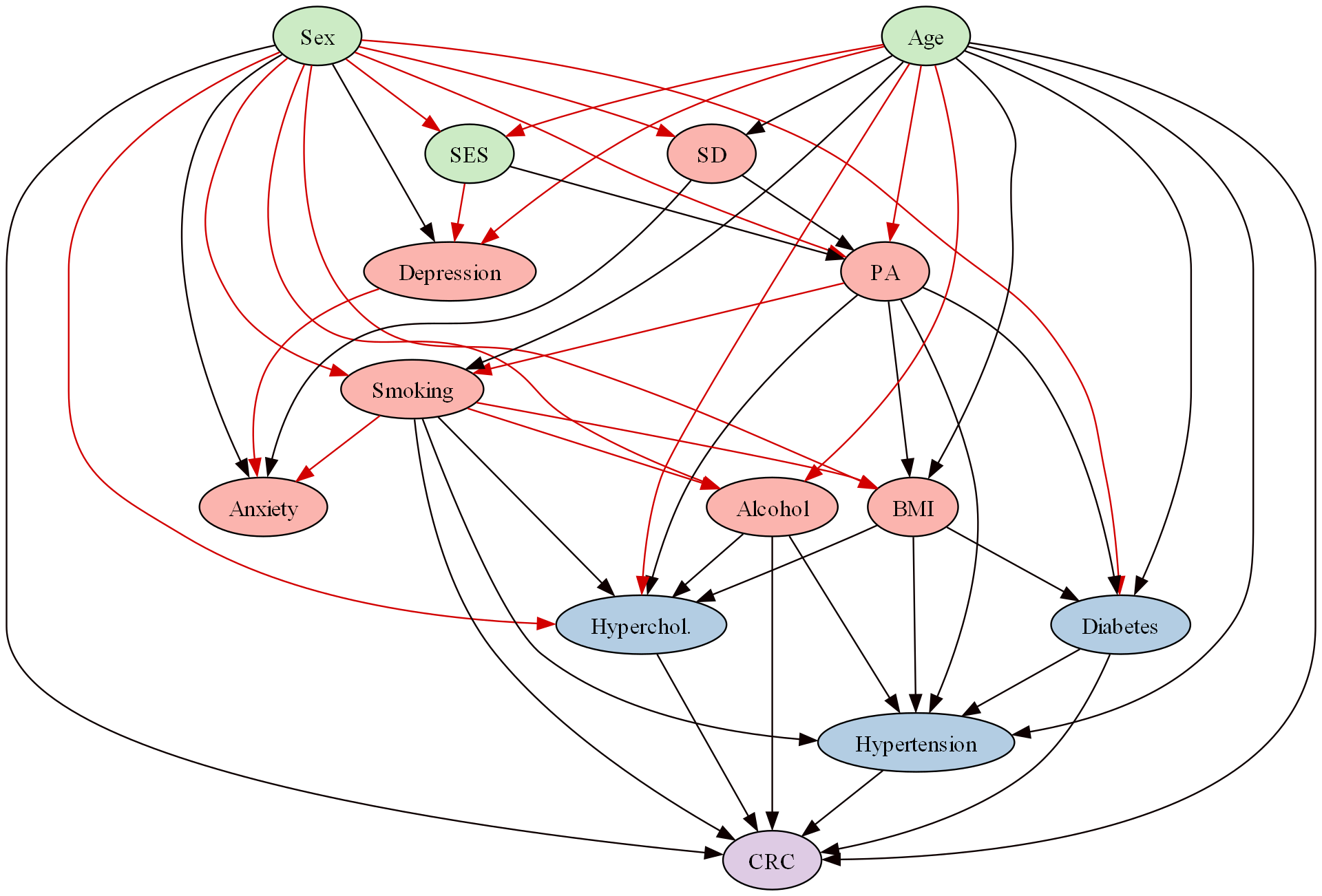}
\caption{Originating CRC Bayesian Network \cite{corrales2024colorectal}}
 \label{bn}
\end{figure}

Our model relies 
  on a BN predictive model 
  previously developed for CRC risk mapping purposes \citep{corrales2024colorectal}
  . The BN aggregates exhaustive expert information and data from a large occupational health assessment study. It relates modifiable (physical activity, sleep duration, alcohol consumption, smoking status, body mass index (BMI), anxiety, depression) and non-modifiable (sex, age, socio-economic status) risk factors as well as medical conditions (hypercholesterolemia, hypertension, diabetes) relevant to CRC, assessing such relations through local probability distributions at the nodes. Figure \ref{bn} presents the original BN, with variables in different colors depending on whether they refer to non-modifiable (green) or modifiable (red) risk factors, medical conditions (blue), or the CRC target (purple) variable. Black arrows were initially elicited from expert information, whereas red arrows were discovered and incorporated using 
   a large database of annual occupational health assessments (2M people). Procedures to assess the probability tables and the BN use for risk mapping and influential variable detection are 
 detailed in \cite{corrales2024colorectal}.

\subsubsection{CRC screening decision model}
 For the construction of the ID for the CRC screening decision-support problem, a set of variables (chance, decision, and values) are defined and incorporated, together with the corresponding arcs, into the original BN.
  
\subsubsubsection*{Hypotheses}\label{subsubsection:hypotheses}

Before presenting the variables, let us first introduce four hypotheses used in constructing the model. 

\begin{enumerate}
    \item Rather than the prevalent {\em one-size-fits-all} strategies \cite{KASTRINOS}, essentially based on age, we aim to provide more personalized advice using influential variables whose information does not require more than a general practitioner (GP) checkup. In any case, the model is easily adapted if additional information based on other relevant risk factors or medical conditions is available, as Section \ref{individual_cases} will illustrate. 
    \item We adapt to the standard strategy based on applying or not a screening device, say FIT, and, if screened 
    and positive, apply a colonoscopy. The objective of the decision model is to suggest to a person with certain features the most convenient screening policy.
    \item The decision to be made should be based on 
    the selected variables,
    the information provided by the screening results, 
    the costs, the entailed complications, and the patient's comfort. 
    Note that, in doing this, we focus on the short-term outcomes
    of the screening intervention, ignoring pointers to longer-term outcomes like, e.g., expected QALYs gained. This is motivated mostly by our interest in benchmarking novel screening devices for which little information will be typically available. The model allows for assessing the relative importance of various criteria as described below through their integration into a multi-attribute utility function.
    \item Suggested decisions will be based on the maximum expected utility (EU) principle \citep{french}, with the utility vision of either the patient, the doctor, or the health policy maker, 
    as later discussed.
\end{enumerate}
These hypotheses are broadly aligned with the recommendations from the European Network for Health Technology Assessment \cite{eunethta2016there}, reflecting a shared direction while allowing for nuanced differences. Based on them, we describe the remaining elements required to structure qualitatively the problem.

\subsubsubsection*{New nodes and arcs}

To complete the design of the ID, the following nodes 
and arcs are added to the BN in Figure \ref{bn}

\paragraph{Decision nodes.}

The {\em screening method} implemented is considered as a decision variable. Its potential alternatives are the currently most 
common CRC screening methods, to wit:
{\em gFOBT}, 
 {\em FIT}, 
 {\em blood-test}, {\em stool DNA test}, {\em computed tomography colonography} (CTC) and {\em colon capsule} (CC). Of course, we also include the possibility of conducting {\em no screening}.

 A second decision node refers to the possibility of {\em performing a colonoscopy}. It is a successor of the previous one and, hence, we consider the possibility of performing it after (or without) screening. Note that it is standard in many European countries to perform a colonoscopy if screening suggests the presence of CRC, say through a positive FIT. On the other hand, several countries opt for directly performing a colonoscopy on susceptible patients. Both possibilities are thus covered within our model and, even enriched, as they can be combined, and further information, beyond age, is used to support the corresponding screening decisions. 

\paragraph{Chance nodes.}

Besides the chance nodes from the original BN, we include three additional ones. First, we consider the potential {\em complications} associated with the eventual screening and colonoscopy interventions \cite{barre2020cost}, which are {\em bleeding}, {\em retention}, {\em perforation}, {\em death}  and, obviously, {\em no complications}. The other two nodes refer to the results of both interventions, {\em screening results} and {\em colonoscopy results}, assimilated to two possible reports: {\em predicted true} (interpreted as screening or colonoscopy suggesting the presence of CRC) and {\em predicted false} (interpreted otherwise). We include as well a {\em  No result } state to handle the case when the corresponding intervention is not actually performed.

\paragraph{Value nodes.}
We introduce a multiple criteria preference model \cite{french} to support CRC screening decisions. As primary criteria, the model includes the {\em cost of complications}, the {\em cost of the intervention}, the {\em comfort of the intervention} and the {\em information provided by screening and/or colonoscopy}. Finally, we include a value node aggregating the four criteria through a multi-attribute utility function.

\paragraph{Arcs}
Arcs from the original BN are preserved. Besides, we include arcs which essentially reflect our starting hypothesis and the information relations between decisions and values.

First of all, based on hypothesis 1, initially we assume that the screening decision is made knowing variables that are easy to obtain or request (BMI, Age, PA, Sex, Alcohol consumption status, Smoking status, Sleeping duration), hence the arcs from such variables to the screening node. Notwithstanding these, the diagram is easily modifiable if we know other variables from the patient, like its eventual hypertension, as Section \ref{section:3} illustrates. 

The arc between the decision nodes and the arc connecting the screening results with the colonoscopy decision reflect hypothesis 2 and facilitate covering all standard screening protocols. In turn, the arcs going into the value nodes reflect hypotheses 3 and 4. Note that some arcs are displayed in a lighter color. This is because the value function defined in Section \ref{section:preference_models} will depend on the parent nodes of CRC and the decision nodes. However to simplify the analysis and visual structure of the model we just kept the original color of the arcs coming from results of screening, results of colonoscopy, and CRC.

\subsubsection{Final influence diagram }

\begin{figure}[h]
\centering
\includegraphics[width=0.9\textwidth]{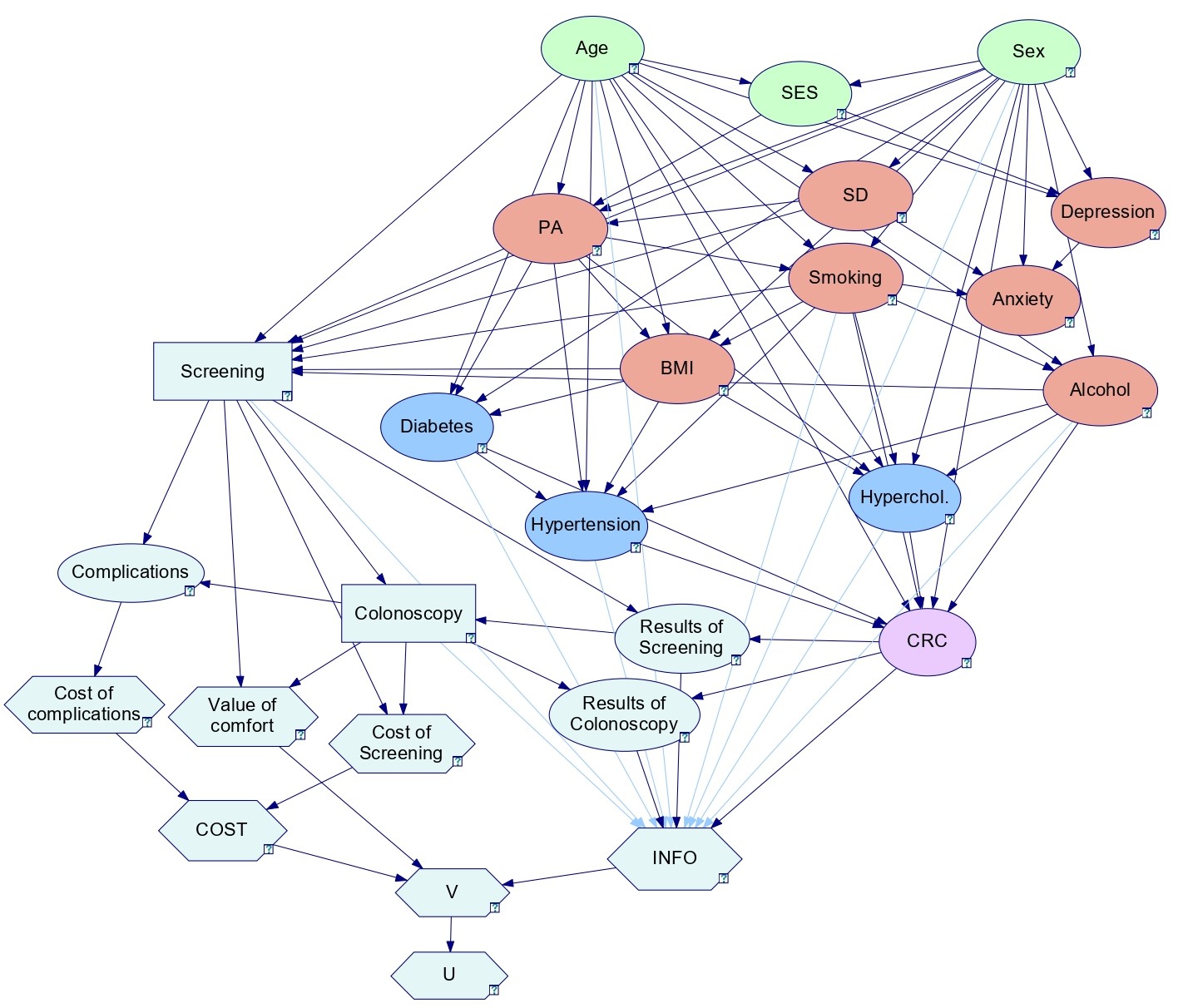}
\caption{CRC Screening Decisions Influence Diagram}
\label{id}
\end{figure}

Figure \ref{id} reflects the structure of the ID produced. In line with our comments, the possible influential variables considered for the screening decision include those whose retrieval does not require further than a GP checkup. However, the socioeconomic situation together with the presence of depression and anxiety have not been taken into consideration as optimizing expected utility based on these variables generates obvious ethical conflicts. The diagram was presented to several medical experts who validated it for concept and meaning.


\subsection{Quantifying the CRC screening decision model}

This section describes the probability and preference models included in the proposed ID for CRC screening decision support.

\subsubsection{Probability models at chance nodes}

First, the ID inherits the probability models at nodes from the original BN. We refer to \cite{corrales2024colorectal}  
for details concerning how the corresponding tables were built based on prior distributions and the available occupational health assessment database. Note that, as a consequence, the probabilities available were assessed with some uncertainty through posterior predictive distributions. We shall use here just the mode of such distributions.

For the other three chance nodes, we used public sources, mainly \cite{barre2020cost}, to obtain the required parameters. In particular, Table \ref{tab:sens_spec} (Appendix \ref{appendix:prob}) includes the  sensitivities and specificities of screening methods and that of colonoscopy, from which we deduce the required probabilities in the nodes {\em results of screening} and {\em results of colonoscopy}, given the presence (or not) of CRC and the method chosen.  As an example, the probability table at node {\em result of colonoscopy}, which depends on the presence of CRC and the use of colonoscopy would be Table \ref{kk:wed}, with e.g.\ the probability of reporting {\em Doesn't have CRC}, given that the person goes through a colonoscopy and she does not have CRC, being 0.03.
\begin{table}[ht]
    \centering
    \begin{tabular}{|l|c|c|c|c|}
    \hline
        Colonoscopy & \multicolumn{2}{c|}{Yes}  & \multicolumn{2}{c|}{No}   \\ \hline
         CRC    & Yes CRC  & No CRC & Yes CRC & No CRC  \\ \hline
  Has CRC &  0.97  & 0.01 &  0   & 0   \\ \hline
  Does not have CRC & 0.03  &  0.99  & 0   &  0    \\ \hline
  No result &    0 & 0   &  1 & 1    \\ \hline
    \end{tabular}
    \caption{Probabilities of colonoscopy results, given antecessors.}
    \label{kk:wed}
\end{table}

Similarly, the probabilities in the first seven columns of Table \ref{tab:complications}  (Appendix \ref{appendix:prob}) were used to build the probability table of {\em complications} associated with each screening method and colonoscopy. Full details of the tables are available in the accompanying software \cite{Corrales-Decision-Model-Screening-CRC-2024}.
 
\subsubsection{Preference models}\label{section:preference_models}

We discuss first the preference model 
for each criterion and aggregate 
them through a multi-attribute utility model.

\subsubsubsection*{Single criterion preferences}
\paragraph{Cost $c_{int}$ of intervention.}
This criterion aggregates the costs of the 
eventual screening and colonoscopy 
interventions. Both should be minimized. We adopt the costs in 
€ assessed in \cite{barre2020cost} for France, 
displayed in Table \ref{tab:sens_spec} (bottom row).

\paragraph{Cost $c_{comp}$ of complications.}
 These are
 assessed through their entailed
 expected costs in € 
 and obtained from public sources
 in \cite{barre2020cost} for France, available in Table \ref{tab:complications} (right column).

\paragraph{Comfort $com$.}
 Lacking a natural attribute 
 to assess intervention comfort, we used a  {\em constructed attribute } \cite{keeney2005selecting} with four decreasing
 levels, the best level (4)
 referring to {\em no screening}; the worst one (1), 
 corresponding to {\em  colonoscopy}.
 Though, in principle, we could assess this in a personalized manner, we constructed the scale reflected in Table \ref{tab:comfort}, 
 later validated by several experts and patients.

\begin{table}[ht]
    \centering
    \begin{tabular}{|c|p{10cm}|p{3cm}|}
    \hline
    $com$  &   Description &  Interventions \\ \hline
    4 &  The patient does not experience any discomfort &  No screening  \\ \hline
    3 &  The patient experiences a minor discomfort 
    or the test implies a small inconvenience: time lost, emotional difficulty, or slight physical pain. & FIT, gFOBT, sDNA, Blood-test       \\ \hline
    2  & The discomfort experienced by the patient is noticeable. There is a noteworthy emotional aversion and a few moments of physical discomfort.
      &    CTC, CC \\ \hline
    1  & The discomfort is very significant. The test causes some periods of pain resulting in remarkable distress. & Colonoscopy  \\  \hline  
    \end{tabular}
    \caption{Comfort levels for interventions.}
    \label{tab:comfort}
\end{table}

 \noindent \citet{torrance1996multiattribute} provide scales within their HUI:2 quality of life system covering several criteria with one of them, 
  pain, close to ours. However, their effort is addressed towards assessing the quality of life over the years by creating a utility function, whereas ours, in line with a 
  short-term health assessment focus \cite{eunethta2016there}, is more geared towards just avoiding uncomfortable episodes.

Importantly, when both screening and colonoscopy 
  are implemented in a patient, the comfort value would be that of 
  colonoscopy, that is 1,  since this is perceived as 
  much more uncomfortable than 
  any of the six analyzed screening methods.
 
\paragraph{Information $v_{\textit{info}}$ provided.}

The more complex attribute
  refers to short-term
informational effects of interventions. These
essentially entail moving from a state of uncertainty based on the probability of a person having CRC (and that of not having it) given their features and the same probability (and its complementary) when we know, as well, the screening and/or colonoscopy results. 

As an example, consider the case of a male adult, age  44-54, with normal sleep duration, physically active, normal weight, non-smoker, and with low alcohol consumption. Further, the patient is negative in all medical conditions considered. We shall use it as a \emph{benchmark} patient below. Table \ref{tab:probs_ref_patient} displays the CRC probability for this individual and his eventual probabilities after screening with FIT and colonoscopy for various results. For instance, a FIT with positive result would move the individual from the prior uncertainty $(0.0009, 0.9991)$ to the posterior $(0.02, 0.98)$.  We would like to assess the information provided by various results taking into account such probabilities (and the complementary ones of not having CRC).
\begin{table}[ht]
    \centering
    \footnotesize
        \begin{tabular}{|c|c|c|c|c|} 
        \hline
        $p(CRC)$ & $p(CRC | FIT-)$ & $p(CRC | FIT+)$ & $p(CRC | FIT+ , COL-)$ & $p(CRC | FIT+ , COL+ )$ \\ 
        \hline
        0.0009  & 0.0002 & 0.02 & 0.0006 & 0.65 \\ 
        \hline
    \end{tabular}
    \caption{Probabilities of CRC depending on screening outcomes for benchmark patient.}
    \label{tab:probs_ref_patient}
\end{table}

\noindent Therefore, we seek ways to evaluate such changes in uncertainty, that is to assess the value of the information that the interventions provide concerning the presence of CRC. 

Assume that we want to assess the amount of uncertainty reduced by a screening strategy. Let us denote by $CRC$ the random variable describing the presence of CRC, and $R_s$ and $R_c$ the respective variables describing the results of screening and colonoscopy. Our interest is in calculating $MI(CRC; (R_s, R_c))$ where $MI$ designates the \emph{mutual information (MI)} function \cite{cover1999elements}. Note that both $R_s$ and $R_c$ depend on the decision of which screening test to use and whether or not to perform a colonoscopy, while the distribution of $CRC$ will depend on the evidence collected from its parent nodes in the ID. However, we shall not make explicit such dependence to lighten the notation. Then, the 
   mutual information is written as 
\begin{align} \label{formula:mut_info_2}
    MI(CRC; (R_s, R_c)) & = MI(CRC; R_s) + MI((CRC; R_c) | R_s) \\ \notag
    & = \mathbb{E}_{crc, r_s} \left[ \log \left( \frac{p(crc, r_s)}{p(crc)p(r_s)} \right) \right] +  \mathbb{E}_{crc, r_s, r_c} \left[ \log \left( \frac{p(crc, r_c | r_s)}{p(crc | r_s) p(r_c|r_s)} \right) \right] \\  \notag
    & = \mathbb{E}_{crc, r_s, r_c} \left[ \log \left( \frac{p(crc| r_s)}{p(crc)} \right)  + \log \left( \frac{p(crc| r_s, r_c)}{p(crc | r_s)} \right) \right] .
\end{align}
 Now, as the suggested decision in our problem will be based on maximum expected utility (hypothesis 4), a natural value function that will effectively quantify the information provided by a screening strategy would be
\begin{align*}
    \textit{pmi}(crc, r_s , r_c) = \log \left( \frac{p(crc| r_s)}{p(crc)} \right)  + \log \left( \frac{p(crc| r_s, r_c)}{p(crc | r_s)} \right).
\end{align*}
We refer to this value as the \emph{pointwise mutual information} of $CRC$ and $(R_s, R_c)$. Although the domain of this function is the set of real numbers, the \textit{MI} (that is, the expectation of the \textit{PMI}) will always be positive and bounded by the entropy of either of the variables.  
As we are interested in the uncertainty concerning the presence of $CRC$, we shall normalize the \textit{MI} dividing it by its entropy $H(CRC) = - \sum p(crc) \log{p(crc)}$, so that the final value function lies in the interval $[0,1]$. In summary, the value of information $v_{\textit{info}}$ provided by a screening strategy concerning the presence of CRC will be given by the \emph{normalized} or \emph{relative pointwise mutual information (RPMI)} defined through 
\begin{align}
    v_{\textit{info}}(crc, r_s, r_c) = \textit{pmi}(crc, r_s , r_c) \   \big/ \  H(CRC).
\end{align}
Intuitively, the expected value of this function 
 refers to the proportion of uncertainty reduced by the 
 screening strategy from the total uncertainty in relation 
  to the presence of $CRC$. 
  
  Importantly, this proposal 
    facilitates acknowledging some difficulties usually encountered 
    in the preference elicitation scenarios considered in the CRC domain. For instance, in relation to the psychological cost of unnecessary assessments 
    associated with false positives \cite{love1981value} and the possibility of a delay in the CRC diagnosis and subsequent dissuasion of participants for later assessment \cite{ferlizza2021roadmap}, the proposed mechanism weighs these situations through their corresponding uncertainty.

As examples, Tables \ref{tab:definition_info_no_screening_ref_pat} and  \ref{tab:definition_info_FIT_ref_pat} respectively provide the $v_{\textit{info}}$ for our benchmark patient when not undertaking screening and when performing FIT.
 Positive values correspond to correct predictions; negative values, to wrong ones; and, finally, zero indicates no change in uncertainty. Values corresponding to cases in which CRC is present are larger in magnitude as they have a low probability. This conforms with the argument that a missed CRC positive scenario is much worse than a misdiagnosed healthy patient \cite{ioannidis2011false}. 

\begin{table}[h]  
    \centering
    \resizebox{0.6\textwidth}{!}{%
    \begin{tabular}{|c|c|c|c|c|c|c|}
    \hline
    Scr & \multicolumn{6}{c|}{No screening}  \\ \hline
    $R_s$ & \multicolumn{6}{c|}{No pred}   \\ \hline
    CRC  & \multicolumn{3}{c|}{False} & \multicolumn{3}{c|}{True} \\ \hline
    Col & No Col & \multicolumn{2}{c|}{Colonoscopy} & No Col & \multicolumn{2}{c|}{Colonoscopy}  \\ \hline
    $R_c$ & No pred & Pred False & Pred True & No pred & Pred False & Pred True \\ \hline
    $v_{\textit{info}}$ & 0.0 & 0.12	& -11.44 & 0.0 & -509.19 & 654.93 \\ \hline
    \end{tabular}
    }
    \caption{No screening $v_{\textit{info}}$ for benchmark.}
    \label{tab:definition_info_no_screening_ref_pat}
\end{table}

\begin{table}[h]  
    \centering
    \resizebox{\textwidth}{!}{%
    \begin{tabular}{|c|c|c|c|c|c|c|c|c|c|c|c|c|}
    \hline
    Scr &  \multicolumn{12}{c|}{FIT} \\ \hline
    $R_s$ & \multicolumn{6}{c|}{Pred False}   & \multicolumn{6}{c|}{Pred True} \\ \hline
    CRC  & \multicolumn{3}{c|}{False} & \multicolumn{3}{c|}{True} & \multicolumn{3}{c|}{False} & \multicolumn{3}{c|}{True} \\ \hline
    Col & No Col & \multicolumn{2}{c|}{Colonoscopy} & No Col & \multicolumn{2}{c|}{Colonoscopy} & No Col & \multicolumn{2}{c|}{Colonoscopy} & No Col & \multicolumn{2}{c|}{Colonoscopy}  \\ \hline
    $R_c$ & No pred & Pred False & Pred True & No pred & Pred False & Pred True & No pred & Pred False & Pred True & No pred & Pred False & Pred True \\ \hline
    $v_{\textit{info}}$ & 0.09 & 0.12	& -2.97 & -196.80 & -706.08 & 466.52 & -2.59 & 0.04 &	-151.00 & 448.05 & -58.63	& 966.01 \\ \hline
    \end{tabular}%
    }
    \caption{FIT $v_{\textit{info}}$ for benchmark.}
    \label{tab:definition_info_FIT_ref_pat}
  \end{table}
\noindent Recall that we look for the best screening strategy and contemplate the decision to perform the colonoscopy depending on the output of the first screening. Thus, the value tables will not contain the expected $v_{\textit{info}}$ but rather the extension of its values depending on the results of screening, that is, $\mathbb{E}_{crc, r_c | r_s}[v_{\textit{info}}]$, which does not necessarily lie in the $[0,1]$ interval, as Table \ref{tab:expected_info_screening_ref_pat} portrays. Indeed, observe how for the first alternative regarding no screening, the value lies in $[0,1]$ as no other result is possible and, thus, coincides with its expectation. When screening is performed, not much information is gained when the result predicts {\em false}, the most likely scenario in general. However, the information provided is much larger when the prediction is {\em true}, increasing its value when a colonoscopy is further performed. Notice how in a low prevalence scenario highly specific tests, like FIT, provide more information, whereas highly sensitive tests, like sDNA,  do not have as much of an impact.

\begin{table}[h]  
    \centering
    \resizebox{0.8\textwidth}{!}{%
    \begin{tabular}{|c|c|c|c|c|c|c|c|c|c|c|}
    \hline
    Screening & \multicolumn{2}{c|}{No screening} & \multicolumn{4}{c|}{FIT} & \multicolumn{4}{c|}{sDNA} \\ \hline
    Result of Scr & \multicolumn{2}{c|}{No pred}  & \multicolumn{2}{c|}{Predicted False} & \multicolumn{2}{c|}{Predicted True} & \multicolumn{2}{c|}{Predicted False} & \multicolumn{2}{c|}{Predicted True} \\ \hline
    Colonoscopy & No Col & Col & No Col & Col & No Col & Col & No Col & Col & No Col & Col \\ \hline
    Exp. $v_{\textit{info}}$ & 0.0 & 0.532 & 0.049 &	0.187	& 5.722 &	15.802 & 0.086 & 0.134 & 0.911 & 4.394
 \\ \hline
    \end{tabular}%
    }
    \caption{Expected information of different policies for reference patient}
    \label{tab:expected_info_screening_ref_pat}
\end{table}

\begin{figure}[ht]%
    \centering
    \includegraphics[width=0.65\textwidth]{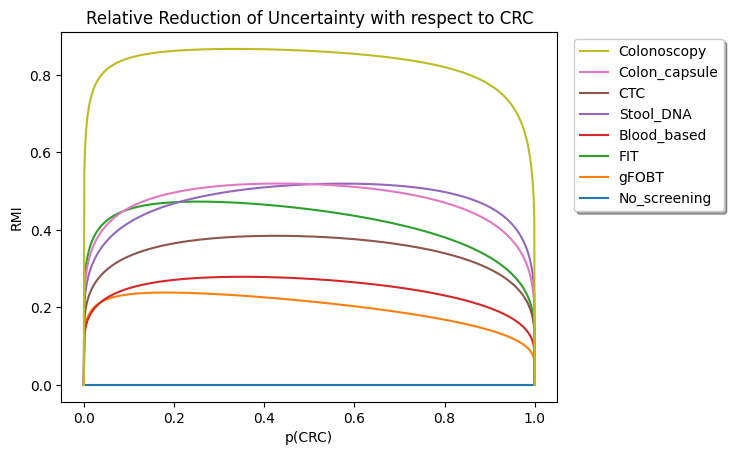}%
    \caption{$v_{\textit{info}}$ for various CRC screening alternatives.}%
    \label{fig:rel_cond_mut_info}
\end{figure}

Figure \ref{fig:rel_cond_mut_info} plots the $v_{\textit{info}}$ function for all possible values of the probability of having CRC facilitating comparison of screening methods in terms of information. Observe that blood-test, CTC, and gFOBT are bounded above by the rest of screening methods in terms of the information measure used.  From this, we conclude that we could discard blood-test and CTC as they are equal or worse to FIT in the four criteria \textit{(information, comfort, specificity, sensitivity)} considered. This is not the case, however, for gFOBT as it is the cheapest method and has non-dominated performance metrics. Further observe that at very low probabilities of CRC, FIT is the method that provides the best information on its own.
 
Note that the asymmetries of \emph{$v_{\textit{info}}$} in Figure \ref{fig:rel_cond_mut_info} derive from the differences between specificity and sensitivity. More specific tests, like FIT, take their maximum at lower $p(CRC)$ values, whereas more sensitive ones, like sDNA, do so at larger $p(CRC)$ values.

\subsubsubsection*{Multiple criteria preference aggregation}

Lastly, let us aggregate the criteria through a utility function $u$, first 
 employing a multicriteria value function, then transformed to take into account risk aversion, see 
 \cite{gonzalez2018utility} for a detailed conceptual 
 description. 

We first aggregate the costs in € so that  $cost= c_{int} + c_{comp}$ for each intervention. The information provided by the screening strategy is already in compact form and, thus, we next define the global value function as a weighted aggregation of the intervention costs and information, taking into account comfort. To wit, for a fixed comfort level $k$, under reasonably general conditions \citep{gonzalez2018utility}, we use a general weighted additive value model. However, after initial numerical experiments with the original information and cost scales, we decided to $\log_{10}$ this last one, adopting the model
\begin{align*}
    & v(cost, \textit{info}, \textit{comf} = k) = \lambda_k \times v_{\textit{info}} - \log_{10}(cost + 1),
\end{align*}
where $\lambda_k$ is a weighting factor that depends on the comfort of the screening strategy and serves as a trade-off between information and log-cost. 

The elicitation of the $\lambda_k$ parameters is delicate and we adopted the following strategy. Assume that only one test is performed at each time and the amount of information corresponds to the reduction of uncertainty through a single independent test, that is, $v_{info} = MI(CRC, R_s) / H(CRC)$. Consider two screening methods with the same comfort level $k$, yet different costs and values, say $(\textit{info}_1, \textit{cost}_1)$ and $(\textit{info}_2, \textit{cost}_2)$, where, typically, the more informative the method is, the more expensive it will be.  Assume that no method dominates the other, in the sense of being both cheaper and more informative than the other, and that the individual declaring his preferences reveals that he favours $(\textit{info}_1, \textit{cost}_1)$ to $(\textit{info}_2, \textit{cost}_2)$, represented as $(\textit{info}_1, \textit{cost}_1) \succ (\textit{info}_2, \textit{cost}_2)$.  Then, we interactively ask the individual for a  $\overline{\textit{cost}}$ value  smaller than $ \textit{cost}_2 $ such that $(\textit{info}_1, \textit{cost}_1) \sim (\textit{info}_2, \overline{\textit{cost}})$. Such options would have the same value, that is,
\begin{align}\label{info_cost_equality}
     \lambda_k \times \textit{info}_1 - \log_{10}(\textit{cost}_1+1) = 
     \lambda_k \times \textit{info}_2  - \log_{10}(\overline{\textit{cost}}+ 1),
\end{align}
and we would solve for $\lambda_k=  ( \log_{10}((\textit{cost}_1 + 1) / (\overline{\textit{cost}}+1))/ ( \textit{info}_1 - \textit{info}_2 ) $. Such value would then be subject to standard consistency checks, see e.g. \cite{french1986decision}. A more robust estimation of the parameter would perform this exercise for each pair of screening methods at each comfort level, obtain the corresponding estimations, and reconcile and robustify them through their median. This final estimation would typically be more robust than just a single estimation made out of a chosen pair of screening methods, but has the drawback of requiring a larger elicitation effort on behalf of the decision respondent. Note that we would expect that for lower discomfort, the value of $\lambda$ should be larger, that is, $\lambda_4 > \lambda_3 > \lambda_2 > \lambda_1$, suggesting that information is more valuable when obtained from a more comfortable screening tool.

\begin{example}
     For reference purposes, consider the benchmark patient to elicit the comfort parameters.  Table \ref{tab:info_ref_patient} presents the $v_{\textit{info}}$ provided by all interventions.
\begin{table}[ht]
    \centering
    \footnotesize
        \begin{tabular}{|c|c|c|c|c|c|c|c|c|} 
        \hline
        & No scr. & gFOBT  & FIT  & Blood & sDNA  & CTC & CC & Colonos. \\ \hline
        
        $v_{\textit{info}}$ & 0 & 0.129 & 0.245 & 0.121 & 0.197 & 0.159 & 0.225 & 0.532 \\ \hline
    \end{tabular}
    \caption{Expected information of interventions benchmark}
    \label{tab:info_ref_patient}
\end{table}
\noindent Figure \ref{cost_info_methods} contains the cost and information provided by the methods for the benchmark patient as well as their comfort. Note that, in this case, the screening alternatives that are non-dominated are {\em no screening}, {\em gFOBT}, {\em FIT}, and {\em colonoscopy}. This will not always be the case as our information value 
 assesses uncertainty and depends on the probability of having CRC,
 see Section \ref{section:sens_analysis} for examples.

\begin{figure}[ht]
\centering
\includegraphics[width=0.55\textwidth]{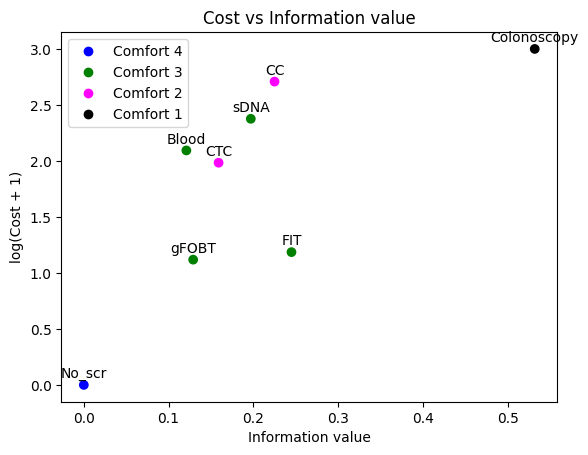}
\caption{Scatterplot}
 \label{cost_info_methods}
\end{figure}

\noindent To illustrate the elicitation process, let us focus on comfort level 2. CTC and CC respectively provide a $v_{\textit{info}}$ of 0.159 and 0.225, while their aggregated costs are 96.51€ and 510.64€. Suppose the individual declares preferring CTC to CC. He is interactively asked how much would the cost of CC have to be so that they are indifferent to one or the other. Assuming that the interview converges to 180€, using (\ref{info_cost_equality})
we obtain $\lambda _2=4.17$. For comfort level 3, FIT and gFOBT respectively provide a $v_{\textit{info}}$  of 0.245 and 0.129, while their aggregated costs are 14.34€ and 12.14€. Suppose the patient declares preferring FIT and that the indifference cost for gFOBT is 3€. Then, we obtain $\lambda _3=5.04$. Repeating this process for all pairs of methods at this
  comfort level and taking their median, we obtain   $\lambda_3 = 6.80$. Table \ref{tab:elicit_lambda} (Appendix \ref{appendix:elicitation}) reflects the full elicitation exercise.
\begin{table}[ht]
    \centering
    \begin{tabular}{|c|c|}
    \hline
     Parameter    & Value  \\ \hline
       $\lambda _1$ &   4.01   \\
      $\lambda _2$ &  4.17   \\
      $\lambda _3$ &    6.80  \\
      $\lambda _4$ &  7  \\ \hline
         \end{tabular}
    \caption{Values of $\lambda$ parameters.}
    \label{value3}
\end{table}

Comfort levels 1 and 4 are slightly different as we only have one option for them. For comfort level 1 we consider a synthetic test,
 providing a $v_{\textit{info}}$ of 0.4 and an indifference-associated cost with respect to colonoscopy of 300. In that case $\lambda _1=4.01$. For comfort level 4, in this case both the $v_{\textit{info}}$ and the cost are 0. Therefore, $\lambda_4$ has no impact in the calculation of the value function, and we just assign $\lambda _4$ so that the monotonicity of the $\lambda$'s is preserved, leading to Table \ref{value3}. 

\hfill $\triangle$

\end{example}

Once the value function is elicited, 
 assuming (constant absolute) risk aversion \cite{gonzalez2018utility} we adopt the following expression for the utility function
\begin{align*}
    u(cost, \textit{info}, \textit{comf} ) =  a  - b \times \exp{ \left( -\rho v(cost, \textit{info}, \textit{comf}) \right) },
\end{align*}
\noindent where $\rho$ is the risk aversion coefficient and $a$ and $b$ are scaling constants constraining the utility to the $[0,1]$ 
interval for the three reference alternatives, determined using, e.g., the classic probability equivalent (PE) method \citep{farquhar1984state}.
\begin{example}{(Cont.)}
Observe that, in our problem, we assume that the presence of CRC is 
 uncertain and, hence, the best outcome in terms of information collected will be detecting a high-risk patient with a very specific test and then performing a colonoscopy.
In turn, the worst outcome in terms of information would be not doing anything as it has no added value. Concerning cost, the best possible option would be a costless test, while the worst possible option would be a very expensive test with all possible complications, having an added cost for the patient. Thus, we
choose as references the pairs ($\textit{cost}^*=0$€, $\textit{info}^* = 15.75$) and ($\textit{cost}_{*} =8131.71$€, $\textit{info}_{*}= 0$).
  
Suppose now that the reference to assess the risk aversion coefficient is ($\textit{cost}=50$€, $\textit{info}=4.1$).\footnote{These values have been chosen to maximize the tool's capability for classification as a validation mechanism, maximizing the F1-score. See the end of Section \ref{section:designing}} Assume all interventions are comfort level 3 and suppose we interactively obtain from the interview that the PE is 0.7. 
 We then have the system 
 \begin{align*}
 \centering
  \begin{cases}
    a  - b \times \exp{ \left( -\rho v(8131.71, 0, 3 ) \right) }=0 \\
    a  - b \times \exp{ \left( -\rho v(0, 15.75, 3 ) \right) } = 1, \\
    a  - b \times \exp{ \left( -\rho v(50, 4.1, 3 ) \right) }=0.7  \\
 \end{cases}
 \end{align*}
\noindent leading to the parameters in Table \ref{utility}.
\begin{table}[ht]
    \centering
    \begin{tabular}{|c|c|}
    \hline
     Param.    & Value  \\ \hline
       $a $ & 1.015     \\
      $ b $ & 0.872    \\
           $\rho $ & 0.039    \\  \hline
    \end{tabular}
    \caption{Utility parameters.}
    \label{utility}
\end{table}
\end{example}  
\hfill $\triangle$

\noindent We shall employ such parameters in our discussion of use cases in Section \ref{section:3}. 


\section{Results}\label{section:3}
Once with our ID built and validated, we
illustrate its application in relevant use cases referring to
providing personalized screening advice, assessing national age-based screening 
strategies, designing personalized national 
screening strategies, and benchmarking novel screening technologies.

\subsection{Managing individual cases}\label{individual_cases}

Let us exemplify a few cases of individuals for which our model would propose different screening advice showcasing how our approach personalizes 
screening recommendations, as summarised in Table \ref{tab:indiv_cases}.

Start with the benchmark patient from Section \ref{section:preference_models} whose probability of having CRC was $0.00085$. Running our model, the decision with maximum EU for him would be not performing screening (EU 0.143). As a comparison, the decision with the second highest EU (0.142) is FIT (followed by a colonoscopy if the predicted result is positive).

Suppose now that the preference model is more risk-seeking, reducing the risk aversion coefficient $\rho$ from 0.039 to $0.005$. Then, the optimal suggested policy would be FIT (EU $0.147$), followed by a colonoscopy if the FIT result is positive 
 (and no colonoscopy if negative). Furthermore, all of the tests, except for CC and blood-based, are preferred to no screening. Intuitively, in the context of this problem, a risk-seeker would be willing to make a larger expense for the information.

As mentioned, the model is easily adaptable when additional medical information is available, beyond the seven variables obtainable through a GP visit. This is done by just adding the corresponding arrows that go from the new variables of interest to the decision node. As an example, suppose that we also know that the benchmark is hypertense and has diabetes. His probability of having CRC is much larger ($0.0039$). Using the model, the advice would be sDNA (EU $0.146$), followed by a colonoscopy if the prediction is positive.

Suppose now that for a given person we have access to an exogenous variable that changes the probability of CRC to 0.1, 
 e.g.\ based on knowing that the patient has CRC family antecedents. In this situation, the recommendation for the first decision would be FIT with an EU of $0.183$; the policy with the highest EU would be FIT followed by no colonoscopy even if the result is positive.
  The reason behind this is that the change in uncertainty that a test can produce at these relatively high levels of CRC probability is not 
  worth the cost of a colonoscopy. Indeed, the information provided by FIT in this case is already quite significant as it changes the prior probability of CRC from $p(CRC) = 0.1$ to $p(CRC|FIT+) = 0.710$ when the result of FIT is positive, rendering somewhat redundant the results of a colonoscopy.

 \begin{table}[ht]
    \centering
    \begin{tabular}{|l|l|l|l|l|l|l|l|l|}
    \hline
         & 1st & EU & 2nd & EU & \multicolumn{2}{l|}{EU Pred False} & \multicolumn{2}{l|}{EU Pred True} \\ 
         & recom & & recom & & No col & Col & No col & Col  \\
         \hline
        Benchmark patient & No scr & .143 & FIT & .142 & - & - & - & -  \\ \hline
        BP risk seeking & FIT & .147 & sDNA & .145 &  \textbf{.139} & .133 & .293 & \textbf{.370} \\ \hline
        BP added ev. & sDNA & .146 & FIT  &  .145 & \textbf{.084} & .056 & .289 &   \textbf{.536 }  \\ \hline
        Exogen. var. $p=.1$ & FIT & .183 & sDNA & .173 & \textbf{.131} & .098 & \textbf{.631} & .554 \\
        \hline
        
    \end{tabular}
    \caption{Personalized screening strategies in the four cases}
    \label{tab:indiv_cases}
\end{table}

\subsection{Assessing the French national screening strategy}\label{section:assessing}

Our goal now is to assess, as an example, the current \emph{one-size-fits-all} French screening strategy based on age. More precisely, the target population for CRC screening are citizens in the age range [50-74] which do not have any hereditary condition or familiar CRC antecedents; when FIT is positive, participants are suggested to undertake a colonoscopy \cite{pellat2018results}. Let us present three cases that are not properly prioritized 
in the current strategy.
\begin{enumerate}
    \item Consider a man with age [54-64], normal sleep, normal BMI, physically active, non-smoker, low alcohol consumption and not having diabetes or hypertension. His probability of having CRC is $0.0022$. Because of his age, he would be required to undertake FIT. However, the proposed model suggests that it is better to administer sDNA (EU $0.145$) rather than FIT (EU $0.144$). Both would be followed by a colonoscopy if the screening prediction is CRC-positive.
    
    \item Consider now a man with age [44-54], with similar characteristics but having diabetes and hypertension. His probability of having CRC is $0.0039$. Because of his age, he would not be called to participate in the screening program. However, the model suggests this patient should undertake sDNA (EU $0.146$), followed by colonoscopy if sDNA is positive.

    \item Finally, for the case of a man with age [44-54], overweight, normal sleep, high-alcohol consumption, physically inactive, and ex-smoker, the probability of having CRC will be $0.0018$. The proposed model 
    suggests that the maximum EU method is FIT with $0.143$, followed by a colonoscopy if its result is positive.
\end{enumerate}
These examples show how the current age-based strategy may fail in various ways at detecting cases with highest EU. As an example, the second patient should be prioritized over the first one within a screening program. Further, sDNA may be more recommendable for higher-risk patients, while FIT is a better choice for moderate-risk patients.  Predictive models for decision-making can aid in redistributing the efforts for more efficient screening programs, as we next show.

\subsection{Designing a national personalized screening strategy}\label{section:designing}

Let us discuss now how the design of a national screening strategy could be based on our 
 model. For this, we use a database with around $350000$ individuals 
  whose records were kept for testing purposes for the BN model in Section \ref{section:2.1}, as fully described in \cite{corrales2024colorectal}.
 We use all thirteen available variables in the network in this use case.
   As our database does not include data from screening or colonoscopy results, we simulate them based on the sensitivity and specificity information of various interventions available in Table \ref{tab:sens_spec}.

When designing a screening program, we should expect constraints on the maximum number of colonoscopies and screening operations performable because of lab, personnel, and device availability, as well as due to budget limits. The issue is, then, how do we allocate such resources using the decision support model in Section \ref{section:2}. Our approach will assign screening methods in order of maximum EU: individuals with higher EU will be offered screening earlier as they are assimilated with the population benefiting more from screening. The intervention choice will be that providing highest EU to the individual; once we saturate the $n_i$ tests available for the $i$-th screening method, we remove it from the list of available methods and search for those of maximum EU among the remaining ones. This process continues until we reach all available test limits or cover the entire targeted population. We assume that there is no limit on the available colonoscopies after a positive screening prediction as these are fundamental for a correct diagnosis.This is a reasonable assumption as it is the standard health practice.

Let us see how this strategy performs by comparing three setups: 
\begin{enumerate} 
\item We assume there are no constraints when applying this new strategy.
\item Constraints are set on the number of tests for each screening method, as expressed in Table \ref{tab:screening_counts} middle row. Specifically, we limit the maximum number of direct colonoscopies to 3000; gFOBT to 30000; FIT to 42000; blood-based tests to 7000; sDNA to 6000; CTC to 2000; and, CC to 2000, therefore being able to cover 91000 individuals (out of the 350000). 
\item Finally, the current national age-based strategy. 
\end{enumerate}

With the parameters developed throughout the text, the distribution of recommended tests and results for the first two strategies would be as Table \ref{tab:screening_counts} shows, where the first row indicates tests administered when no constraints are included; the middle row indicates test constraints; and, finally, the last row indicates the number of tests when constraints are included. Interestingly, only sDNA and FIT are the recommended strategies. The excess of recommended sDNA tests is distributed by administering FIT and implementing no screening. No direct colonoscopy would be recommended in this general context. For 291707 patients, no screening is preferred to the remaining alternative screening methods even when setting operational limits.

\begin{table}[!ht]
    \centering
    \footnotesize
    \begin{tabular}{|l|l||l|l|l|l|l|l|l||l|}
    \hline
        ~ & Nothing & Colon. & gFOBT & FIT & Blood & sDNA & CTC & CC & Total \\ \hline
        No lim recom. & 291323 & 0 & 0 & 560 & 0 & 47824 & 0 & 0 & 48384\\ \hline
        Op. limit & $\infty$ & 3000 & 30000 & 42000 & 7000 & 6000 & 2000 & 2000 &  - \\ \hline
        Final recom. & 291707 & 0 & 0 & 42000 & 0 & 6000 & 0 & 0 & 48000 \\ \hline
        National &  290633 & 0 & 0 & 49074 & 0 & 0 & 0 & 0 & 49074 \\ \hline
    \end{tabular}
    \caption{Comparison of strategies with no constraints (top row) and with constraints (bottom row). Operational limits are shown in the middle row.}
    \label{tab:screening_counts}
\end{table}

We apply the three strategies described and obtain the results for 200 simulations. Tables \ref{fig:old_strat_classification}-\ref{fig:new_strat_w_lim_classification} contain their confusion matrices for CRC classification with values corresponding to the mean over the 200 simulations and their standard deviations. Table \ref{fig:old_strat_classification} shows how the current age-based strategy has a poor performance in detecting patients with CRC, that is a poor sensitivity (0.36), while the proposed unconstrained new strategy, shown in Table \ref{fig:new_strat_classification}, increases detection sensitivity to 0.45. However, this comes with the trade-off of a decrease in precision (0.82 to 0.64), resulting in a larger number of false positives for the proposed strategy. Concerning cost, the unconstrained new strategy entails an increase in the average cost ($44.62$ € vs. $7.24$€ per patient). To account for that excessive increase in cost and the considerable precision decrease, we set operational limits (middle row Table \ref{tab:screening_counts}) on the number of tests usable for each method. Table \ref{fig:new_strat_w_lim_classification} shows the corresponding classification metrics from this strategy, with more moderate costs ($12.79$ € vs. $7.24$€ per patient). There is still an increase in the sensitivity (0.37) and a decrease in precision (0.79), but both are more subtle now. Depending on the cost-information trade-off and risk attitude, the take on the results of the new strategy is that such investment may be recommended to reach higher sensitivity levels and thus increase the number of detected patients in screening programs. However, this is not always manageable and demands an operational limit in more realistic scenarios.

\begin{table}[h!]
    \centering
    \begin{tabular}{|c|c|c|}
        \hline
        & \textbf{Predicted No CRC} & \textbf{Predicted CRC} \\
        \hline
        \textbf{No CRC} & $339472.1 \pm 4.0$ & $16.9 \pm 4.0$ \\
        \hline
        \textbf{CRC} & $139.8 \pm 4.4$ & $78.3 \pm 4.4$ \\
        \hline
    \end{tabular}
    \caption{Mean classification results for current strategy. Cost per patient: 7.24€}
    \label{fig:old_strat_classification}
\end{table}

\begin{table}[h!]
    \centering
    \begin{tabular}{|c|c|c|}
        \hline
        & \textbf{Predicted No CRC} & \textbf{Predicted CRC} \\
        \hline
        \textbf{No CRC} & $339425.2 \pm 8.2$ & $63.8 \pm 8.2$ \\
        \hline
        \textbf{CRC} & $121.3 \pm 3.1$ & $96.7\pm 3.1$ \\
        \hline
    \end{tabular}
    \caption{Mean classification results for new strategy, no constraints. Cost per patient: 44.62€}
    \label{fig:new_strat_classification}
\end{table}

\begin{table}[h!]
    \centering
    \begin{tabular}{|c|c|c|}
        \hline
        & \textbf{Predicted No CRC} & \textbf{Predicted CRC} \\
        \hline
        \textbf{No CRC} & $339466.5 \pm 4.9$ & $22.5 \pm 4.9$ \\
        \hline
        \textbf{CRC} & $134.8 \pm 4.1$ & $83.2 \pm 4.1$ \\
        \hline
    \end{tabular}
    \caption{Mean classification results for new strategy with operational constraints. Cost per patient: 12.81€}
    \label{fig:new_strat_w_lim_classification}
\end{table}

For validation purposes, assume that the current system can cover 49,074 FIT tests, the number of people in the population from our 2016 dataset in the age range [54,64]. Assuming this, we compare the current strategy with our model by taking the 49,074 patients with highest FIT EU and performing 200 simulations to assess the differences in performance and cost. Table \ref{fig:new_strat_w_lim_classification_FIT} shows a subtle increase in sensitivity when using the model, correctly detecting an average of one more patient (an increase of $1\% $) and reducing the cost of the whole strategy by around 33000€. Although the improvement might seem relatively low, when extrapolated to a real-sized population, the method can save many lives and money. As an example, France and Spain 
  respectively have populations of $8,591,286$ and $6,583,183$ within the screening age [54-64]. Thus, just an implementation of the model within the mentioned target population would detect around 175 and 134 more positive patients in France and Spain, respectively.

\begin{table}[h!]
    \centering
    \begin{tabular}{|c|c|c|}
        \hline
        & \textbf{Predicted No CRC} & \textbf{Predicted CRC} \\
        \hline
        \textbf{No CRC} & $339472.5 \pm 4.0$ & $16.54 \pm 4.0$ \\
        \hline
        \textbf{CRC} & $138.9 \pm 4.6$ & $79.1 \pm 4.6$ \\
        \hline
    \end{tabular}
    \caption{Mean classification results for new strategy on 49074 patients with highest FIT utility. Cost per patient: 7.14€}
    \label{fig:new_strat_w_lim_classification_FIT}
\end{table}

\subsection{Benchmarking of new screening devices}\label{section:benchmark}

Given the increasing importance of CRC from a public health policy perspective, it is likely that, in the near future, there will be novel CRC screening devices.\footnote{As an example, this is one of the aims of the ONCOSCREEN project 
\href{https://oncoscreen.health/}{https://oncoscreen.health/}}
  We discuss here how the approach proposed may be used  to benchmark new devices. For illustration purposes, suppose we have come out with 
   two new devices with features as in Table \ref{pedobarna}.
\begin{table}[h]  
    \centering
        \begin{tabular}{|c|c|c|c|c|}
    \hline
    Device & Cost & Specif. & Sensit. & Comfort \\ \hline
    Dev1  &   250 & 0.85 & 0.8 & 2 \\
    Dev2  &     3 & 0.85 & 0.94 & 3 \\ \hline
    \end{tabular}%
        \caption{Features of two new screening devices, Dev1 and Dev2}
    \label{pedobarna}
\end{table}

\begin{figure}[ht]
\centering
\includegraphics[width=0.6\textwidth]{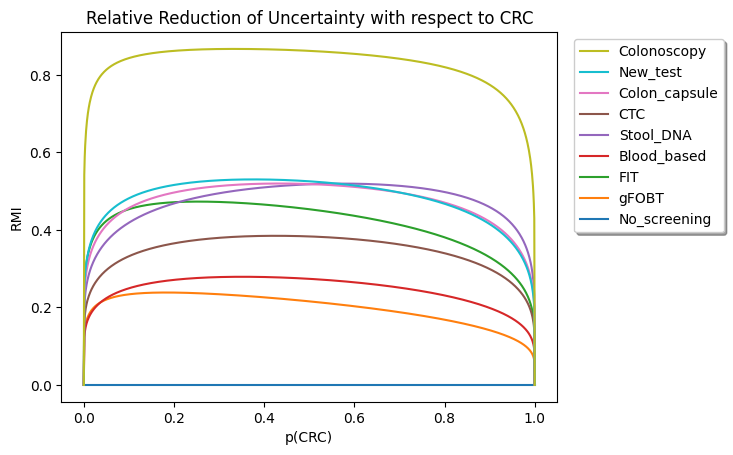}
\caption{Plot of $v_{\textit{info}}$ at different levels of $p(CRC)$}
 \label{new_test}
\end{figure}

First of all, we should check whether the new devices are not dominated by the current ones. As an example, we would reject Dev1 because it is dominated by sDNA, whose features are $\textit{cost} = 236.88$€, $\textit{specif} = 0.866$, $\textit{sensit} = 0.923$ and $\textit{comfort} = 3$.  However, Dev2 is non-dominated.
Let us assess it with our model. Figure \ref{new_test} shows that for most of the range of the CRC probability below 0.55, the new test reduces uncertainty more than currently available screening tests. For extremely low CRC probability values, FIT still provides more information due to its 
  better specificity. However, as Dev2 is much cheaper than FIT and has the same comfort value, we have that for the benchmark patient, the EU of the new test is $0.179$, which is higher than that of the
  previous recommendation which was {\em no screening} (Section 3.1).

\begin{table}[!ht]
    \centering
    \footnotesize
    \begin{tabular}{|l|l||l|l|l|l|l|l|l|l||l|}
    \hline
        ~ & Nothing & Colon. & gFOBT & FIT & Blood & sDNA & CTC & CC & New & Total\\ \hline
        No lim. recom. & 0 & 0 & 0 & 0 & 0 & 0 & 0 & 0 & 339707 & 339707 \\ \hline
        Op. limit & $\infty$ & 3000 & 30000 & 42000 & 7000 & 5000 & 2000 & 2000 & 50000 & - \\ \hline
        Final recom. & 287920 & 0 & 0 & 1787 & 0 & 0 & 0 & 0 & 50000 & 51787 \\ \hline
    \end{tabular}
    \caption{Comparison of strategies when adding the new test.}
    \label{fig:counts_new_test} 
\end{table}
Suppose that we implement this test. As Table \ref{fig:counts_new_test} shows, with the considerations that our model makes in terms of cost, comfort, and information, we would be recommending this test to the entire population. However, this is in general hard to achieve and would entail a large cost (63.86€ per patient) due to the high number of colonoscopies carried out. Hence, we generate a case in which we have 50,000 new Dev2 tests and the same resources as in the examples from Section \ref{section:designing}. Table \ref{fig:classification_new_test_w_lim} provides the classification results with this strategy, which essentially increases the number of detected patients by more than 10, an increase of around $20\%$. The number of false positives also increases, as well as the total cost (9.85€ per patient), as more patients are being predicted CRC positive and require a colonoscopy. However, the increment in the F1 score, a common classification measure that balances the importance of true and false positives and negatives \cite{van1979information}, from 0.50 in the original strategy to 0.54 shows that the benefit would be significant in classification.

\begin{table}[h!]
    \centering
    \begin{tabular}{|c|c|c|}
        \hline
        & \textbf{Predicted No CRC} & \textbf{Predicted CRC} \\
        \hline
        \textbf{No CRC} & $339458.5 \pm 5.4$ & $30.5 \pm 5.4$ \\
        \hline
        \textbf{CRC} & $126.8 \pm 3.8$ & $91.2 \pm 3.8$ \\
        \hline
    \end{tabular}
    \caption{New strategy including new tests. Cost per patient: 9.85€}
    \label{fig:classification_new_test_w_lim}
\end{table}

\subsection{Sensitivity analysis} \label{section:sens_analysis}

We conclude by performing several relevant sensitivity analysis with our model, specifically in relation to the utility function obtained.

\subsubsection{Sensitivity analysis with respect to the probability equivalent}\label{section:sens_analysis_PE_method}

\begin{figure}[ht]%
    \centering
    \includegraphics[width=0.9\textwidth]{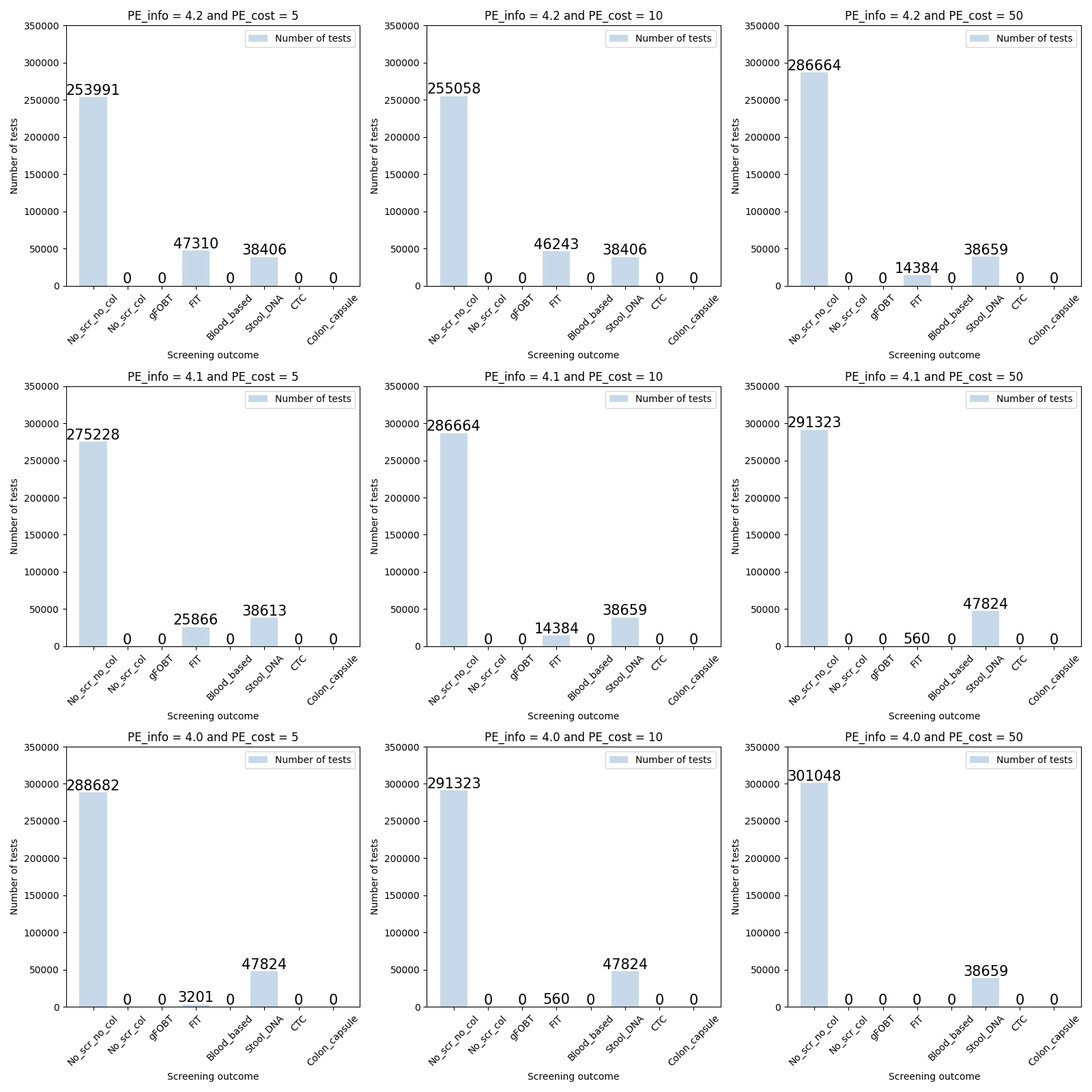}%
    \caption{Plot with different PE info and PE cost}%
    \label{fig:sens_analysis_screening_counts}
\end{figure}

We first analyze the impact of the probability equivalent (PE) elicited to determine the risk aversion coefficient in the allocation of screening tests.
Figure \ref{fig:sens_analysis_screening_counts} plots the optimal screening resource allocations showing how information generally impacts test allocation more than cost, as changes are more noticeable vertically than horizontally. In general, the total number of recommended tests increases as the weight given to the value of information increases and decreases as the weight given to cost increases.

The intuition behind this is that sDNA is, given the comfort levels, the recommended test for higher-risk patients as it is the most sensitive one and thus provides the highest quality information. However, suppose a larger weight is given to information at lower costs. In that case, the number of FITs will increase as it is one of the cheapest methods and provides quite competitive information regarding CRC presence. Suppose now that information is valued more at higher costs. In that case, the number of sDNA tests allocated will increase as long as its information-cost tradeoff surpasses the utility of not performing screening. 

\subsubsection{Sensitivity analysis with respect to comfort parameters}\label{sec:sens_analysis_elicitation}

Let us discuss now sensitivity with respect to the comfort parameters $\lambda$. Remember that by repeating the elicitation exercise for comfort 3 screening methods, we ensured some robustness of the method. However, for the other comfort levels, this cannot be done due to the absence of additional tests for comparison. Moreover, the elicitation of comfort values depends on the available tests and does not account for alternative or future tests and their respective information values. This is complicates extrapolating results to new strategies, as seen in Section \ref{section:benchmark} where, with the established elicited parameters using the original tests, the new test is recommended to the entire population due to its great features, being this recommendation far from manageable. 
We look into how the variability in these values can lead to differences in recommendations and thus highlight the importance of consistency and robustness in the elicitation protocol.

\begin{figure}[h]%
    \centering
    \subfloat[\centering ]{{\includegraphics[width=0.5\textwidth]{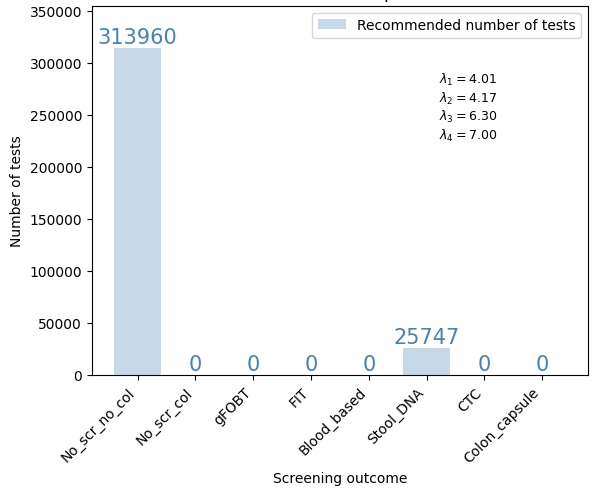} }}%
    \subfloat[\centering ]{{\includegraphics[width=0.5\textwidth]{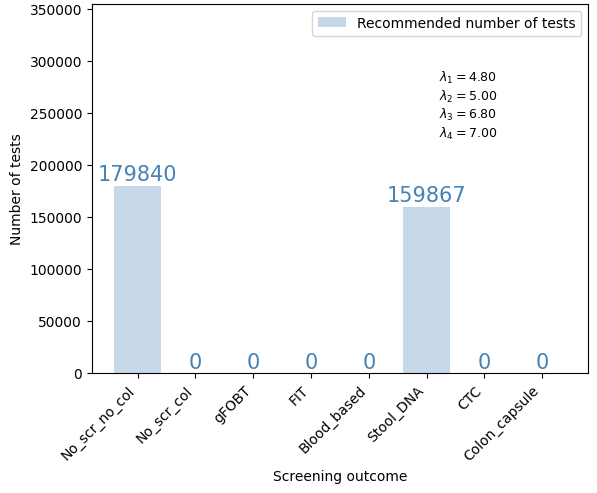} }}%
    \caption{Screening count distribution for different $\lambda$ values}%
    \label{fig:sens_analysis_elicitation}
\end{figure}

Figure \ref{fig:sens_analysis_elicitation} shows two screening test recommendation distributions. The first one corresponds to decreasing the value of level 3 comfort to $\lambda_3 = 6.3$; observe how a lower comfort parameter reduces the number of screenings performed, increasing the amount of non-screened people. The second case corresponds to raising the value of level-1 comfort to $\lambda_1 = 4.8$ and level 2 to $\lambda_2 = 5$; the increase in the weight given to the comfort value for colonoscopy has a large effect in increasing the number of recommended sDNA tests. Recall that a colonoscopy is usually performed after a positive screening result,  and thus, as the weight given to the comfort value for colonoscopy increases, the expected value of all screening methods also increases. However, this change is more noticeable for sDNA, it being the most sensitive tool, thus detecting more CRC positive cases.

Therefore, a moderate modification of comfort parameters impacts program results. To ensure consistency,  we would recommend following the elicitation protocol in a moderate number of patients chosen at random, asking them about their indifference between cost, information, and comfort. Repetition and randomization would enable a more robust estimation of comfort parameters which would benefit model applicability.

\section{Discussion}

We have developed a decision analysis model and decision support system for personalised screening in connection with CRC detection. Stemming from an earlier predictive model \cite{corrales2024colorectal}
we incorporated new chance, decision, and value nodes as well as new arcs, assessing the new probability tables. We introduced a multi-attribute utility model to support what combination of screening and colonoscopy decisions should be implemented for an individual with certain features. This could help to strengthen the monitoring systems of colorectal screening programmes, as has been proposed for European countries \cite{altobelli2019differences}. As showcased, this facilitates a more personalized approach to CRC screening as well as designing large-scale screening strategies taking into account constraints on treatment availability. Apart from the viability of the use of the model in public health management, it can also serve to simulate screening scenarios, regarding how to distribute screening tests according to different risk levels and estimate the costs of different strategies. The system enables as well benchmarking of novel screening devices. 

The preference model incorporated is the one discussed in Section \ref{section:preference_models}, and assimilated to that of the policy makers designing or assessing a screening strategy. However, the preference model may be easily modified, and eventually individualised, to promote a more personalized approach. in line with \cite{volk2018guideline}.

As illustrated in Section \ref{individual_cases}, the model may be adapted and enhanced when additional variables are taken into account. In particular, this may be the case when data concerning diet and genetics is available, which was not the case in our initial dataset. The availability of information on additional variables can be very useful in improving the possibilities of better personalising screening programmes. In any case, the major strength of this work lies in its development of a practical framework and methodology which is adaptable to more complex scenarios.

At least, two future lines emerge at least from this work. The first one refers to benchmarking the new screening devices that are being developed within the ONCOSCREEN project with the methodology specified in Section \ref{section:benchmark}. The second one, given the low acceptance of CRC screening programs in many countries, will focus on amplifying our model with incentive mechanisms to promote screening adoption.

\bibliography{sample}
\pagebreak 

\appendix
\section{Probabilities available from public sources}\label{appendix:prob}

To assess the probabilities in the chance nodes introduced,
  Table \ref{tab:sens_spec} compiles the performance information of the screening methods as well as their cost with data available from \citet{barre2020cost}, based on French sources.
 
\begin{table}[ht]
    \centering
    \begin{tabular}{|l|l|l|l|l|l|l|l|}
    \hline
        &gFOBT & FIT & BldBsd & sDNA & CTC & CC & Colons. \\ \hline
        Sensitivity & 0.45 & 0.75 & 0.66 & 0.923 & 0.8 & 0.87 & 0.97 \\ \hline
        Specificity & 0.978 & 0.966 & 0.91  & 0.866 & 0.89 & 0.92 & 0.99 \\ \hline
        Cost € & 12.14 & 14.34 & 123.13 & 236.88 & 95.41 & 510.24 & 1000  \\ \hline
    \end{tabular}
    \caption{Specificity, sensitivity, and cost in € of interventions.}
    \label{tab:sens_spec}
\end{table}

\noindent We also require the probabilities and expected costs of
complications as reflected in Table \ref{tab:complications}
based on \cite{barre2020cost}. 
\begin{table}[ht]
    \centering
    \begin{tabular}{|l|l|l|l|l|l|l|l||l|}
    \hline
        & gFOBT & FIT & BlbBsd & sDNA & CTC & CC & Colons & Cost \\ \hline
    
        None & 1 & 1 & 1 & 1 & 0.9996 & 0.9997 & 0.998 & 0 €\\ \hline
        Bleed. & 0 & 0 & 0  & 0 & 0 & 0 & 0.0006 & 1241€\\ \hline
        Reten. & 0 & 0 & 0 & 0 & 0 & 0.0003 & 0 &  1241€ \\ \hline
        Perfor. & 0 & 0 & 0 & 0 & 0.0004 & 0 & 0.001 & 2810€ \\ \hline
        Other & 0 & 0 & 0 & 0 & 0 & 0 & 0.0004 & 6621€ \\ \hline
    \end{tabular}
    \caption{Probabilities and expected costs of complications for
    CRC interventions.}
    \label{tab:complications}
\end{table}


\pagebreak
\clearpage

\section{Full elicitation of comfort parameters} \label{appendix:elicitation}

Table \ref{tab:elicit_lambda} contains full details of the calculations used to assess comfort parameters. 
\begin{table}[h]
    \centering
    \begin{tabular}{|c|c|c|c|c|c|c|c|}
        \hline
        Comfort & Scr. method & Cost & Info & Preference & Indiff. cost & $\hat{\lambda}_k$ & $\lambda_k$ \\
        \hline
        \multirow{2}{*}{1} & Colonos & 1000 & 0.530 &  & --  & \multirow{2}{*}{$\lambda_1 = 4.01$} & \multirow{2}{*}{$\lambda_1 = 4.01$}\\
        \cline{2-6}
         & Synth. & -- & 0.4 & $\times$ & 300€ & & \\
        \Xhline{3\arrayrulewidth}
        \multirow{2}{*}{2} & CTC & 95.41 & 0.159 & $\times$ & -- & \multirow{2}{*}{$\lambda_2 = 4.17$} & \multirow{2}{*}{$\lambda_2 = 4.17$} \\
        \cline{2-6}
         & CC & 510.24 & 0.225 & &  180€ &  & \\
        \Xhline{3\arrayrulewidth}
        \multirow{2}{*}{3} & gFOBT & 12.14 & 0.129 &  & 3€ & \multirow{2}{*}{$\lambda_3 = 5.04$} &  \multirow{12}{*}{$\bar{\lambda}_3 = 6.80$} \\
        \cline{2-6}
         & FIT & 14.34 & 0.245 & $\times$ & -- & & \\
        \cline{2-7}
        \multirow{2}{*}{3}  & gFOBT & 12.14 & 0.128 & $\times$ & -- & \multirow{2}{*}{$\lambda_3 = 10.57$} & \\
        \cline{2-6}
         & Blood test & 125.13 & 0.121 &  & 10€ & & \\
        \cline{2-7}
        \multirow{2}{*}{3} & gFOBT & 12.14 & 0.128 & $\times$ & -- & \multirow{2}{*}{$\lambda_3 = 16.28$} & \\
        \cline{2-6}
         & sDNA & 236.88 & 0.197 &  & 170€ &  & \\
        \cline{2-7}
        \multirow{2}{*}{3} & FIT & 14.34 & 0.244 & $\times$ & -- & \multirow{2}{*}{$\lambda_3 = 6.40$} & \\
        \cline{2-6}
         & Blood test & 125.13 & 0.121  &  & 1.5€ & &\\
        \cline{2-7}
        \multirow{2}{*}{3}  & FIT & 14.34 & 0.244 & $\times$ & -- & \multirow{2}{*}{$\lambda_3 = 7.2$} & \\
        \cline{2-6}
         & sDNA & 236.88 & 0.197 &  & 6€ & & \\
        \cline{2-7}
        \multirow{2}{*}{3}  & Blood test & 125.13 & 0.121 &  & 80€ & \multirow{2}{*}{$\lambda_3 = 6.17$} & \\
        \cline{2-6}
         & sDNA & 236.88 & 0.197 & $\times$ & -- & & \\
        \Xhline{3\arrayrulewidth}
        4 & No scr. & 0 & 0 & -- & -- & -- & $\lambda_4 = 7$ \\
        \Xhline{3\arrayrulewidth}
    \end{tabular}
    \caption{Eliciting parameter $\lambda_k$. X indicates the preferred alternative in the corresponding pairwise comparison.}
    \label{tab:elicit_lambda}
\end{table}

\end{document}